\documentclass[aps,prd,a4paper,superscriptaddress,nofootinbib,10pt, two column]{revtex4-1}
\usepackage{amsmath,amssymb,graphicx,multirow,dcolumn,bm,latexsym,soul,nicefrac}
\usepackage[colorlinks,linkcolor=blue,citecolor=blue,urlcolor=blue]{hyperref}
\newcounter{RomanNumber}
\usepackage[normalem]{ulem}

\allowdisplaybreaks
\newcommand{\be}{\begin{equation}}
\newcommand{\ee}{\end{equation}}
\newcommand{\bea}{\begin{eqnarray}}
\newcommand{\eea}{\end{eqnarray}}

\begin{document}

\title{Black holes in General Relativity and beyond}

\author{Enrico Barausse}
\affiliation{Institut d'Astrophysique de Paris, CNRS \& Sorbonne
 Universit\'es, UMR 7095, 98 bis bd Arago, 75014 Paris, France}
\affiliation{SISSA, Via Bonomea 265, 34136 Trieste, Italy and INFN Sezione di Trieste}
\affiliation{IFPU - Institute for Fundamental Physics of the Universe, Via Beirut 2, 34014 Trieste, Italy}

\begin{abstract}
    The recent detections of gravitational waves from binary systems of black holes are in remarkable agreement with the predictions of General Relativity. In this pedagogical mini-review, I will go through the physics of the different phases of the evolution of black hole binary systems, providing a qualitative physical interpretation of each one of them. I will also briefly describe how these phases would be modified if gravitation were described by a theory extending or deforming General Relativity,
or if the binary components turned out to be more exotic compact objects than black holes.
\end{abstract}
\maketitle

\section{Introduction}
The binary black hole (BH) systems detected by ground based gravitational wave (GW) interferometers (first by the two LIGOs alone~\cite{PhysRevLett.116.061102,Abbott:2016nmj,TheLIGOScientific:2016pea,Abbott:2017vtc,Abbott:2017gyy}, then by the LIGO-Virgo network~\cite{Abbott:2017oio,LIGOScientific:2018mvr}) are now ten, and their signals have been found to be in excellent agreement with the predictions of General Relativity (GR). While GR had previously passed experimental tests with flying colors in 
weak field and/or mildly relativistic regimes (in the solar system~\cite{Will:2014kxa} and in binary pulsars~\cite{Hulse:1974eb,Damour:1991rd}), the LIGO-Virgo detections provide for the first time evidence that GR is viable in the strong gravity, highly relativistic and dynamical regime relevant for BH binaries~\cite{TheLIGOScientific:2016src,TheLIGOScientific:2016pea}. Moreover, they also support the hypothesis that the binary's components, as well as the merger remnant, are really BHs, as opposed to more exotic compact objects~\cite{PhysRevLett.116.061102}.

Nevertheless, as always in physics, these statements on the correctness of GR and on the BH hypothesis come with several caveats, and they are only correct within the ``experimental errors'' affecting the measurements. In other words, while large non-perturbative effects in the GW signal seem to be ruled out by current observations~\cite{TheLIGOScientific:2016src,TheLIGOScientific:2016pea}, deformations of the general relativistic signal that are either sufficiently small and/or in the frequency ranges where LIGO and Virgo are least (or not at all) sensitive may still be in agreement with observations.

The purpose of this note is to briefly scope out where deviations from GR or from the BH hypothesis might still hide in the GW signal from binary systems, while remaining in agreement with observations. I will do so by conventionally splitting the signal in four parts:  an early low-frequency portion produced by the binary's inspiral (Sec.~\ref{insp});   one produced by the merger (Sec.~\ref{merger}) with the subsequent ringdown phase (Sec.~\ref{RD}); and a post-ringdown phase produced by putative exotic near horizon physics (Sec.~\ref{PRD}). (This fourth phase is of  course absent in GR.) I will qualitatively interpret each part by relying on physical intuition from the test particle limit, Newtonian physics and BH perturbation theory.

\section{The inspiral}\label{insp}
At low orbital (and GW) frequencies, i.e. at large separations, the evolution of the binary
is driven by the emission of GWs. The latter backreact on the system by carrying energy and angular
momentum away from it, and thus cause the binary to slowly inspiral to smaller and smaller separations. In GR, this ``(gravitational) radiation reaction'' is well described, in the limit of small relative binary velocities $v\ll c$, by the quadrupole flux formula, according to which the energy flux emitted by a binary with masses $m_1$ and $m_2$ and separation $r$ is given  by~\cite{Peters1964}
 \begin{equation}\label{Edot}
\dot{E}_{\rm quadrupole} = \frac{32 G}{5 c^3} \left(\frac{G m_1 m_2}{r^2}\right)^2
                           \left( \frac{v}{c} \right)^2\,.
\end{equation}
For a quasi-circular binary, Kepler's law yields $v=(G M/r)^{1/2}$ (with $M=m_1+m_2$), and
 the GW frequency $f$ is given by twice the orbital frequency $f_{\rm orb}=v/(2 \pi r)$. One can then
 relate the GW frequency to the binary's total energy $E$ (in the center of mass frame) by
 $E=-G m_1 m_2/(2r)$, and re-express Eq.~\eqref{Edot} as~\cite{Peters1964}
 \begin{equation}\label{eq:dfdt_gw}
     \dot{f}=\frac{96 G^{5/3}}{5 c^5} \pi^{8/3} \mathcal{M}^{5/3}\, f^{11/3}\,,
     \end{equation}
where $\mathcal{M}= M \nu^{3/5}$, with $\nu= m_1 m_2/M^2\leq 1/4$, is the {\it chirp mass}.
A simple lesson to  be drawn from this  equation is that GW observations of the inspiral determine the chirp mass to high accuracy, simply by measuring the frequency and its rate of change. Indeed, the first GW detection (GW150914~\cite{PhysRevLett.116.061102}) determined
the chirp mass of the binary source to be
$\mathcal{M}\approx 30 M_\odot$, which translates into a total mass $M\gtrsim 70 M_\odot$.

The quadrupole formula can be generalized by performing a ``post-Newtonian'' (PN) expansion, i.e. a perturbative expansion of the field equations in the small quantity $v/c\ll1$ (see e.g. Ref.~\cite{blanchet} for a review of the PN formalism). By solving the field equations under that approximation, one can find expressions for the fluxes that reduce to Eq.~\eqref{Edot} in the limit of low frequencies (when $v\ll c$), but which also include subdominant terms in $v/c$ (e.g. the  mass-octupole and current-quadrupole contributions). These terms become important as the binary's separation shrinks. Similar PN-expanded expressions can be found not only for the GW fluxes, but also for the gravitational waveforms and for the Hamiltonian describing the conservative interaction between the two binary components. 

In particular, the PN expansion allows one to account for the effect of BH spins, which appear both in the conservative and dissipative sectors. One remarkable effect of BH spins during the inspiral is given by the modulations of the GW amplitude that they cause when they are misaligned with
the binary's orbital angular momentum~\cite{apostolatos}. Indeed, save for special configurations giving rise to {\it transitional precession}~\cite{apostolatos}, the spins and the orbital angular momentum
undergo {\it simple precession} around the  total angular momentum of the system, whose direction remains approximately constant during the binary's evolution (including even the merger ringdown phase)~\cite{apostolatos,final_spin}. Detecting these spin precession modulations, as well other spin related effects in the gravitational waveforms (e.g. the dependence of the plunge-merger frequency on the spins, which
we will discuss below in Sec.~\ref{merger}, and the enhancement of the GW amplitude caused by large spins aligned with the orbital angular momentum), one can, at least in principle, measure  BH spins. Indeed, in two GW detections (GW151226 and GW170729)
there seems to be evidence in favor of non-zero, probably precessing spins~\cite{Abbott:2016nmj,TheLIGOScientific:2016pea,LIGOScientific:2018mvr}. Much more accurate measurements of BH spins (to within 1--10\% or better) will also become possible with space based detectors such as LISA~\cite{lisa,Klein2016}.

As one moves away from GR to consider more general gravitational theories, the structure of the PN expansion also gets modified. The LIGO-Virgo collaboration has performed phenomenological tests of GR by ``deforming'' the PN expressions for the GW amplitude and phase as functions of frequency. In particular, they allowed for PN coefficients (i.e. coefficients akin e.g. to the $32/5$ appearing in Eq.~\eqref{Edot}) different from the GR predicted values, and tried to constrain the differences away from the latter with observations. This resulted in ``large'' deviations from GR being excluded by the data~\cite{TheLIGOScientific:2016src,TheLIGOScientific:2016pea}.
Note that these deformations of the PN coefficients may effectively  account also for the possibility that 
 the binary components may not be BHs but more exotic objects, which would present different
tidal effects than BHs (see e.g. Ref.~\cite{Cardoso:2017cfl}). 

Another effect that would arise quite naturally from an extension/modification of GR is a low-frequency change of the fluxes and waveforms, resulting e.g. from the existence of BH dipole radiation~\cite{Barausse:2016eii}.
Within GR, gravitational monopole and dipole radiation are forbidden by the conservation of the stress energy tensor, i.e. respectively by the conservation of energy and linear momentum. This is similar to what happens in electromagnetism, where monopole emission is forbidden by the conservation of the electric charge. Suppose now, however, that the theory of gravity describing Nature is not GR (which only possesses a spin-2 graviton, i.e. which only describes gravity by means of a metric), but some more general theory in which gravity is described by a metric {\it and} additional gravitational degrees of freedom (e.g. a scalar field). To avoid the appearance of unwanted fifth forces in existing  experiments, these additional gravitons must be coupled very weakly (if they are coupled at all) to matter, but they may be coupled non-minimally to the metric. As a result, the coupling to matter, which is negligible in weak gravity experiments such as those performed on Earth, may re-appear in  strong gravity systems, where non-linearities of the metric become large enough to mediate an effective coupling between matter and the extra gravitational fields. 

One may parametrize this effective coupling (also known as ``Nordtvedt effect''~\cite{1975ApJ...196L..59E,PhysRev.169.1014,Nordtvedt:1968qs}) by ``charges'' carried by compact objects such as BHs and neutrons stars.\footnote{These charges are instead negligible for stars or objects with weak internal gravity.} These charges, also referred to as ``sensitivities''~\cite{1975ApJ...196L..59E,0264-9381-9-9-015,Will:1989sk,Foster:2007gr,Yagi:2013qpa,Yagi:2013ava,ramos}, can be object-dependent, e.g. they can be different for BHs and neutron stars, or for stars of different compactness (as they need to vanish in the small compactness limit). In this sense, they may be viewed also as parametrizing violations of the universality of free fall (the equivalence principle). The latter holds in GR, but  is generally violated beyond GR, at least in strong field regimes, because of energy and momentum exchanges between matter and the extra  gravitational degrees of freedom, mediated by the large metric perturbations.

Now, just as in Maxwell's theory a binary system of objects carrying unequal electric charges emits dipole radiation, a quasi-circular binary of BHs or neutron stars carrying unequal sensitivities $s_1\neq s_2$ will emit (gravitational) dipole radiation. Such an emission would be enhanced relative to the quadrupole flux of Eq.~\eqref{Edot} by a term $\sim (s_1-s_2)^2 (v/c)^{-2}$, i.e. it would dominate the binary evolution at low frequencies~\cite{1975ApJ...196L..59E,0264-9381-9-9-015,Will:1989sk,Foster:2007gr,Yagi:2013qpa,Yagi:2013ava,ramos}.

Bounds on the existence of this dipole radiation are very strong in binary pulsar systems (see e.g. Ref.~\cite{shao}), but little can be inferred from those bounds about BH dipole emission, since sensitivities can be different for different classes of objects. Indeed, theories exist where the sensitivities are exactly zero for stars (including neutron stars), and non-zero only for BHs~\cite{Barausse:2015wia,Yagi:2015oca}. Bounds on BH dipole emission from LIGO-Virgo are rather loose, since these detectors  cannot observe the low-frequency ($\lesssim 10 $ Hz) signal from the early inspiral because of instrumental limitations (mainly due to seismic and Newtonian noise). However, much more stringent bounds will be made possible by the launch of the ESA-led mission LISA~\cite{lisa}, a space-borne GW detector targeting the mHz frequency band. This will allow for observing the low frequency inspiral of LIGO-Virgo sources before they are detected from the ground~\cite{sesana_multiband}, and also to observe massive BH binaries~\cite{Klein2016} and extreme mass ratio inspirals~\cite{babak17}. This will result in bounds on the BH dipole flux several orders of magnitude stronger than current ones~\cite{Barausse:2016eii}.

\section{The merger}\label{merger}

Under the effect of radiation reaction, a BH binary slowly evolves along quasi-circular orbits with decreasing separation. Unlike in Newtonian mechanics, where circular orbits can exist with arbitrarily small radii around a point mass, in GR there exists an innermost stable circular orbit (ISCO), inside which circular orbits can still exist but are unstable~\cite{Misner:1974qy}. ISCOs do indeed exist for test particles (i.e. in the limit $m_2/m_1\to 0$) around Schwarzschild and Kerr BHs, and they are present also away from the test particle limit, e.g. when one includes next-to-leading order corrections in ${\cal O} (m_2/m_1)$, or when one expands the dynamics of generic mass ratio binaries in a PN series (in the latter case, the ISCO corresponds to the circular orbit with minimum energy, as defined by the PN Hamiltonian mentioned in the previous section)~\cite{letiec,barak,valli}. 

The position of the ISCO, its energy and its angular momentum depend on the parameters of the BH binary, and most notably on the spins. This is easily seen in the test particle limit, i.e. for the ISCO of the Kerr geometry. As the spin of the Kerr BH (projected on the direction of the particle's orbital angular momentum) increases, the ISCO moves inwards, while its energy and angular momentum decrease~\cite{bardeen}. In particular, the ISCO for a test particle moving on retrograde orbits (with respect to the BH rotation) in a Kerr spacetime lies at larger radii than for a particle on prograde orbits~\cite{bardeen}. This behavior is of course a manifestation of the more general ``frame dragging'' effect of GR, whereby matter near BH horizons tends to co-rotate with the BH as seen from infinity. It is also the reason why the radiative efficiency of thin accretion disks, whose inner edge lies at the ISCO of the central BH, is an increasing function of the BH spin, and why their inner edge gets closer and closer to the BH as the spin of the latter increases. Both facts are crucial to estimate the spin of BH candidates by electromagnetic observations such as continuum fitting and iron-K$\alpha$ lines~\cite{Middleton:2015osa,Brenneman:2013oba}. Similarly, any modification of the BH geometry causing an ISCO shift (as a result of a deviation of the gravity theory from GR, or simply because the central object is more exotic than a run-of-the-mill BH) can be constrained, at least in principle, by the same electromagnetic techniques~\cite{MN,Bambi:2011vc}.

As in the test-particle limit, the position of the effective ISCO of a comparable mass binary
moves to smaller separation for larger BH spins~\cite{letiec,valli,hangup}. 
Once the binary reaches the ISCO, it cannot transition to a stable circular orbit and therefore plunges and merges. The plunge-merger phase takes place in a dynamical time $\sim GM/c^3$, i.e. it is much shorter than the inspiral phase. For this reason, the GW fluxes emitted in the plunge-merger phase can be neglected, to first approximation, with respect to those emitted during the inspiral, i.e. most of the energy and angular momentum emitted by a BH binary is lost from large separations down to the effective ISCO of the system. Therefore, as in the case of thin accretion disks (whose radiative efficiency increases with the spin of the central BH), the GW emission efficiency of BH binaries is a growing function of the spins~\cite{morozova} (because larger spins imply smaller ISCO separations, and thus longer inspiral phases and larger integrated GW fluxes).

One can also attempt to estimate the final spin of the BH remnant produced by the merger that follows the plunge by conserving angular momentum. Consider for instance a test particle falling into a BH from a quasicircular equatorial inspiral. Neglecting the fluxes during the plunge-merger, and also neglecting the fraction of the binary's mass emitted in GWs (see Ref.~\cite{kesden} for a treatment that avoids this  second hypothesis), the final spin of the BH is
\begin{equation}
\label{eq:af}
{a}_{\rm fin}=
\frac{1}{(1+q)^2}\left(a_{1} + a_{2} q^2+ \tilde{L}_{\rm ISCO}(a_1) q\right)\,,
\end{equation}
where $q=m_2/m_1\ll1$ and $\tilde{L}_{_{\rm ISCO}}(a_1)$ is the ISCO angular momentum (normalized by $Gm_1 m_2/c$ to make it dimensionless) as a function of the central (``primary'') BH spin $a_1$. Let us look for instance at the simple case of non-spinning BHs, 
 $|a_1|=|a_2|=0$, and extrapolate from the test-particle limit to equal masses, $q=1$. The final predicted spin is then ${a}_{\rm fin}\approx 0.866$, which is quite off compared to ${a}_{\rm fin}\approx 0.686$ predicted by 
 numerical relativity simulations of BH mergers~\cite{caltechcornell}. However, if the primary BH spin $a_1$ used to compute
 $\tilde{L}_{_{\rm ISCO}}$ is replaced by $a_{\rm fin}$, one is left with an algebraic equation in  $a_{\rm fin}$, which can be solved e.g. by iteration~\cite{bkl}.
 Remarkably, this procedure gives an excellent estimate for the final spin, $a_{\rm fin}\approx 0.66$~\cite{bkl}.
 This iterative method can be generalized to generic (even misaligned) spins and mass ratios, and turns out to be always in good agreement with numerical relativity simulations~\cite{bkl,kesden}. (See also Refs.~\cite{final_spin_original,final_spin,hofmann,lousto,leo} for other procedures to estimate the final spin based on similar physical arguments).
 
 At least two lessons can be drawn from this method to estimate the final spin. The first is that there seems to be a mapping between the comparable mass BH binary problem and the dynamics of a test particle in a Kerr spacetime with spin equal to the final spin. The existence of this mapping is hardly surprising because similar mappings between the comparable mass and test-particle two body problems are known to exist, e.g.
 in Newtonian mechanics (where a binary can be mapped into a test particle of mass $m_1 m_2/M$ around a point mass $M=m_1+m_2$), in quantum electrodynamics (where one can map the energy levels of positronium into those of hydrogen~\cite{PhysRevD.1.2349}) and in PN theory itself (with the effective one body  --- EOB --- model, applicable to the inspiral phase of non-spinning~\cite{eob} and spinning~\cite{DJS,eob_spin} BH binaries). Indeed, the success of the iterative method of Ref.~\cite{bkl} to compute the final spin, which we have outlined above, provides justification for extending the EOB model through the merger ringdown phase~\cite{eobIMR}.
 
 Secondly, it should be stressed that if deviations from GR or from the BH paradigm itself were to change the position and angular momentum of the effective ISCO of a BH binary, then Eq.~\eqref{eq:af} and the ``amended'' extrapolation procedure of Ref.~\cite{bkl} would produce a different final BH spin. While direct measurements of the final spin by LIGO and Virgo are currently impossible because of the low signal-to-noise ratio (SNR) of the ringdown~\cite{TheLIGOScientific:2016src}, future detectors such as LISA will allow for measurements of the ringdown with SNR of several hundreds~\cite{ringdown_letter}, thus making direct spin measurements possible. Moreover, even with the current LIGO-Virgo detections one can attempt to perform consistency tests between the inspiral part of the waveforms (which depends on the individual masses and spins) and the plunge-merger-ringdown part (which depends on the final mass and spin). These tests provide at present no signs of large deviations away from GR~\cite{TheLIGOScientific:2016src,TheLIGOScientific:2016pea} (see also the next section).
 
 Another notable constraint on the nature of the components of the LIGO-Virgo binaries can be obtained from simple physical arguments involving the merger part of the waveform. As mentioned previously, the GW150914 BH merger event has chirp mass ${\cal M}\sim 30 M_{\odot}$ ($M\sim 70 M_\odot$), as estimated from the inspiral. The peak of the GW amplitude, corresponding to the merger, is at about $150$ Hz, corresponding to an orbital frequency of $\sim 75$ Hz and a separation (assuming a binary with roughly equal masses) of $\sim350$ km~\cite{PhysRevLett.116.061102}. Since $GM/c^2\sim 200$ km, this means that the components of GW150914 
merge when they are very close (in gravitational radii). If they were more exotic compact objects with sizes $R_{1}+R_{1}\gg GM/c^2$, they would be interacting well before reaching a separation of 350 km, which would trigger an earlier merger than for BHs. This seems to disfavor the hypothesis that the binary components 
could be e.g. boson stars~\cite{PhysRevLett.116.061102}, unless their compactness is very close to that of BHs (which in turns puts restrictions
on the potential of the bosonic degrees of freedom). These bounds are much stronger than those
for BHs in the massive/supermassive range, which come e.g. from tracking S-stars around the massive BH at the center of
our Galaxy~\cite{S2}, and which still allow for a compactness $GM/(R c^2)\gtrsim 10^{-3}$. 
Moreover, note that deviations from GR may also trigger early mergers, especially in the presence of non-perturbative effects, c.f. Ref.~\cite{DS1,DS2} for an example of a theory triggering early mergers of neutron star systems.

\section{The ringdown}\label{RD}
When the two binary components coalesce, the remnant is initially highly perturbed as a result
of the violent merger process. As the system evolves toward an asymptotically stationary state, the
amplitude of these perturbations decreases and eventually becomes sufficiently small for them to
be described by linear theory.
Within GR, linear perturbations $h_{\mu\nu}$ over a generic vacuum spacetime with metric $g_{\mu\nu}$ satisfy
the wave equation~\cite{Poisson_lorenz,Sciama_lorenz}
\begin{equation}\label{eq:box_h}
\Box\, \bar{h}^{\alpha\beta} + 2 R_{\mu\ \nu}^{\ \alpha\ \beta}
        \bar{h}^{\mu\nu} = 
        0\;,
\end{equation}
with $\bar{h}_{\mu\nu}\equiv h_{\mu\nu}-h^\alpha_\alpha\,g_{\mu\nu}/2$
obeying the Lorenz gauge condition $\nabla_\nu \bar{h}^{\mu\nu}=0$, 
and $\Box\equiv g^{\mu\nu}\nabla_\mu\nabla_\nu$.

However, linear perturbations over a Kerr background of gravitational radius $r_s=GM/c^2$ ($M$ being the mass) and spin parameter $a\in[0,1]$ are better described
by the Newman-Penrose scalars~\cite{NP} 
\begin{gather}
{\bf \Psi}_0=-C_{\mu\nu\lambda\sigma}l^\mu m^\nu l^\lambda m^\sigma\,,\\
{\bf \Psi}_4 = -C_{\mu\nu\lambda\sigma}n^\mu m^{*\nu} n^\lambda m^{*\sigma}\,,
\end{gather}
where $C_{\mu\nu\lambda\sigma}$ is the Weyl curvature tensor, and ${\bf l\,,n\,,m\,,m^*}$ constitute a (complex) null tetrad, defined at each spacetime point. 
Note that ${\bf \Psi}_0$ and ${\bf \Psi}_4$ can be thought of as describing ingoing and outgoing GWs. 

Because of the symmetries of the Kerr metric (and in particular the existence of a Killing tensor, besides the Killing vectors associated to stationarity and axisymmetry), the linear perturbation
equations are dramatically simplified by
a decomposition in Fourier modes and 
in spin-weighted {\it spheroidal} harmonics ${}_s S_{lm}$ (with $s=2$ and $s=-2$ for 
${\bf \Psi}_0$ and ${\bf \Psi}_4$), i.e.
by adopting the ansatz~\cite{teuk}
\begin{align}
&\psi(t\,,r\,,\theta\,,\phi) =\nonumber\\&\frac{1}{2\pi}\int e^{-i\omega t}
\sum_{l=|s|}^{\infty}\sum_{m=-l}^{l}
e^{im\phi}\,{}_sS_{lm}(\theta)R_{lm}(r)d\omega\,,
\end{align}
where $\psi$ stands for either 
${\bf \Psi}_0$ or $\rho^{-4}{\bf \Psi}_4$, with $\rho \equiv -1/(r-ia r_s\cos\theta)$.
Indeed, with this decomposition the perturbed field equations reduce to a single ordinary differential equation for the radial function $R_{lm}$~\cite{teuk}:
\begin{equation}\label{psiteuk}
    \Delta \partial^2_r
R_{lm}+2(s+1)(r-r_s)\partial_rR_{lm}+VR_{lm}=0\,,
\end{equation}
with $\Delta=r^2-2 r r_s+a^2 r_s^2$ and
$V$ a (complex) potential depending on $\omega$, $r$, $M$, $a$
$s$, $l$ and $m$. Moreover, by introducing
the tortoise coordinate $r_*$, which
ranges from $-\infty$ at the horizon to
$+\infty$ at spatial infinity, one can zoom in onto the near horizon region, 
and the equation becomes reminiscent of the one-dimensional Schrodinger equation with a potential barrier.

This equation should be solved with boundary conditions corresponding to no radiation incoming from infinity (i.e. the system is isolated) and no radiation exiting the event horizon (since nothing can escape the event horizon in GR). The boundary conditions (ingoing at the event horizon and outgoing at infinity) select
a discrete spectrum of complex frequencies $\omega$ (c.f. the similar problem of the one-dimensional Schrodinger equation with a rectangular potential). The frequencies are known as ``quasi-normal mode'' (QNM) frequencies~\cite{berti_starinets},
and for non-extremal Kerr BHs they have strictly negative imaginary part, i.e. the modes that they describe are damped.
These damped oscillations constitute the ``ringdown'' part of the GW signal from a BH binary.

From the procedure whereby they are derived, which we have just outlined, it is clear that for given multipole numbers $\ell,m$, the QNM frequencies only depend (in GR) on the BH mass and spin. This is in a sense a consequence of the no-hair theorem of GR, which states that BHs can only be described by their mass and spin, and that they do not carry any additional ``charges'' or ``hairs''~\cite{no_hair}. (Note that BHs in GR can carry an electric charge, but for astrophysical BHs that is expected to be zero or negligible~\cite{Barausse:2014tra}.)

The direct measurement of the QNM frequencies and decay times from the LIGO-Virgo detections is  a difficult task because of the low SNR of the ringdown. Indeed, at present one can at most check the compatibility of the ringdown signal with the predictions of GR~\cite{TheLIGOScientific:2016src,TheLIGOScientific:2016pea}, i.e. with the QNM frequencies and decay times that would be expected based on the final remnant spin predicted via Eq.~\eqref{eq:af} (or a similar method),
once the masses and spin measurements from the inspiral part are factored in.

Stronger ringdown signals would allow not only for measuring, directly from the data, the frequency and decay time of the dominant, least damped QNM (the $\ell=m=2$ mode for a quasicircular binary), but also the first subdominant mode (usually the $\ell=m=3$ or $\ell=m=4$, mode depending on the binary's parameters). This would permit performing a null test of the no-hair theorem~\cite{Dreyer:2003bv}. Indeed, as mentioned above, all QNM frequencies and decay times depend only on $M$ and $a$ in GR. Therefore, one could use the $\ell=m=2$ measured frequency and decay time to infer $M$ and $a$, use those to predict the
frequency/decay time of the second QNM, and finally check the agreement with the measured value(s). This kind of test, however, would require not only that the ringdown signal be detectable (i.e. with SNR larger than $\sim 8$), but also
that the SNR be sufficiently large to resolve both the dominant and first subdominant QNM. In practice, this requires an SNR of at least $\sim 90$ in the ringdown phase alone~\cite{Berti:2007zu,ringdown_letter}, which will be achieved routinely with LISA, but which will require
Voyager-type third generation detectors on Earth~\cite{ringdown_letter} (unless one manages to stack several ringdown signals together by rescaling them~\cite{stack}).
 
Several observations can be made at this
stage. First, in theories of gravity extending GR, not only can the spin-2 QNMs differ from GR (as a result e.g. of BHs carrying hairs, or simply because of the different field equations at linear order~\cite{fR}), but there may also be QNMs with different helicities (e.g. scalars, vectors, or even additional tensor modes in bimetric gravitational theories)~\cite{long_review}. These QNMs would give rise to different polarizations in the response of a network of GW detectors, and would be in principle distinguishable from purely spin-2 modes~\cite{eardley_pol}. However, as mentioned previously, non-GR gravitons  must be coupled very weakly to a detector (if coupled at all) to ensure the absence of unwanted fifth forces in experiments testing the (weak) equivalence principle on Earth. Therefore, the coupling of QNMs of helicity $s\neq\pm 2$
to the detector will be strongly suppressed (see eg. Ref.~\cite{DS1} for an explicit example).

Secondly, let us note that in the geometric optics limit (i.e. $\ell\gtrsim m\gg 1$) the equations of linear perturbation theory [Eq.~\eqref{psiteuk}] must reduce to the null geodesics equation. This is most easily seen by inserting the ansatz $\bar{h}_{\mu\nu}\approx A_{\mu\nu} \exp(i S)$ into Eq.~\eqref{eq:box_h}, and expanding in the limit of large $\partial_\mu S$ (i.e.~in the limit of large frequencies and wavenumbers). This gives the null condition $g^{\mu\nu} \partial_\mu S \partial_\nu S=0$,
which coincides with the Hamilton-Jacobi equation for massless particles. From the derivative of this condition, one then obtains the null geodesics equation (c.f. e.g. section 7.8 of Ref~\cite{defelice} for details). This result is of course hardly surprising as it simply amounts to saying that the gravitational wavefronts follow null geodesics (i.e. GWs travel at the speed of light), but it has noteworthy implications for the physics of QNMs. Indeed, since the linear perturbation equations must reduce to the null geodesics equation in the geometric optics limit, the potential $V$ appearing in the QNM master equation \eqref{psiteuk} must reduce to the effective potential governing the motion of null geodesics in Kerr in the limit $\ell\gtrsim m\gg 1$. In particular,
in that limit the position of
the peak of the potential must asymptote to the same location as the (unstable) circular photon orbit, and the frequencies of the QNMs
are simply given by linear combinations of the orbital and frame-dragging precession frequencies of the circular photon orbit~\cite{wkb1,wkb2,wkb3}. (Note
that the frame-dragging precession frequency only appears in the QNM expressions for spinning BHs.) Moreover, one can show that the decay times of the QNMs for $\ell\gtrsim m\gg1$ are related to the Lyapunov exponents of null geodesics near the circular photon orbit, which in turn depend on the curvature of the effective potential for null geodesics near its peak~\cite{wkb1,wkb2}. An intuitive interpretation of these results is that the QNMs are generated at the circular photon orbit and slowly
leak outward, since that orbit is unstable to radial perturbations.

From this point of view, it is clear that the physics of the QNMs is very similar (if not the same) as that of the circular photon orbit, and electromagnetic experiments aiming to explore the latter present clear synergies with GW detectors. One example is given by the Event Horizon Telescope~\cite{eht}, an ongoing very long baseline radio interferometry experiment aiming to observe the near horizon region of Sgr A$^{*}$ (the massive BH at the center of our Galaxy), 
and which has recently provided the first ``shadow image'' of the horizon of the massive BH at the center of M87~\cite{2019ApJ...875L...1E}. Another implication of realizing that QNMs are generated near the circular photon orbit is that BH mimickers that deviate significantly from the Schwarzschild/Kerr geometry in that region may be severely constrained once GW interferometers detect the ringdown part of a binary's signal with sufficiently large SNR.

\section{The post-ringdown}

From the derivation of the QNMs  outlined in the previous section, it is clear that if the boundary conditions at the event horizon are not perfectly ingoing, the QNM frequencies and decay times will in general change. Partially or fully reflective boundary conditions may in fact be physically relevant in case the merger remnant is
not a BH but a more exotic horizonless object (e.g. a wormhole~\cite{Cardoso:2017cfl}, or an ensemble of horizonless quantum states~\cite{Mazur:2004fk,Mathur:2005zp,Mathur:2008nj,Barcelo:2015noa,Danielsson:2017riq,Berthiere:2017tms,Cardoso:2017njb}), perhaps motivated by  attempts to solve the information loss paradox~\cite{Almheiri:2012rt}. Other possible deviations from the GR event horizon paradigm may arise in
 theories that violate Lorentz symmetry in the purely gravitational sector~\cite{Blas:2011ni,ramos}.

In the presence of such deviations from purely ingoing boundary conditions, part of the GW radiation generated at the peak of the potential (i.e. roughly at the circular photon orbit, where QNMs are produced) is reflected off the horizon. Part of this reflected radiation then tunnels through the potential barrier at the circular photon orbit (eventually reaching infinity), while the rest is reflected back again toward the horizon, where the process can restart. Therefore, after the ringdown described in the previous section, which corresponds to radiation generated at the potential's peak and leaking outward, there appears a series of ``echoes''  delayed (relative to the initial ringdown signal) by multiples of twice the light travel time between the circular photon orbit and the effective ``mirror'' at the horizon~\cite{Barausse:2014tra,Barausse:2014pra,Cardoso:2016rao,Cardoso:2016oxy,Cardoso:2017cqb}.

\begin{figure}[thb]
\includegraphics[width=8.7cm]{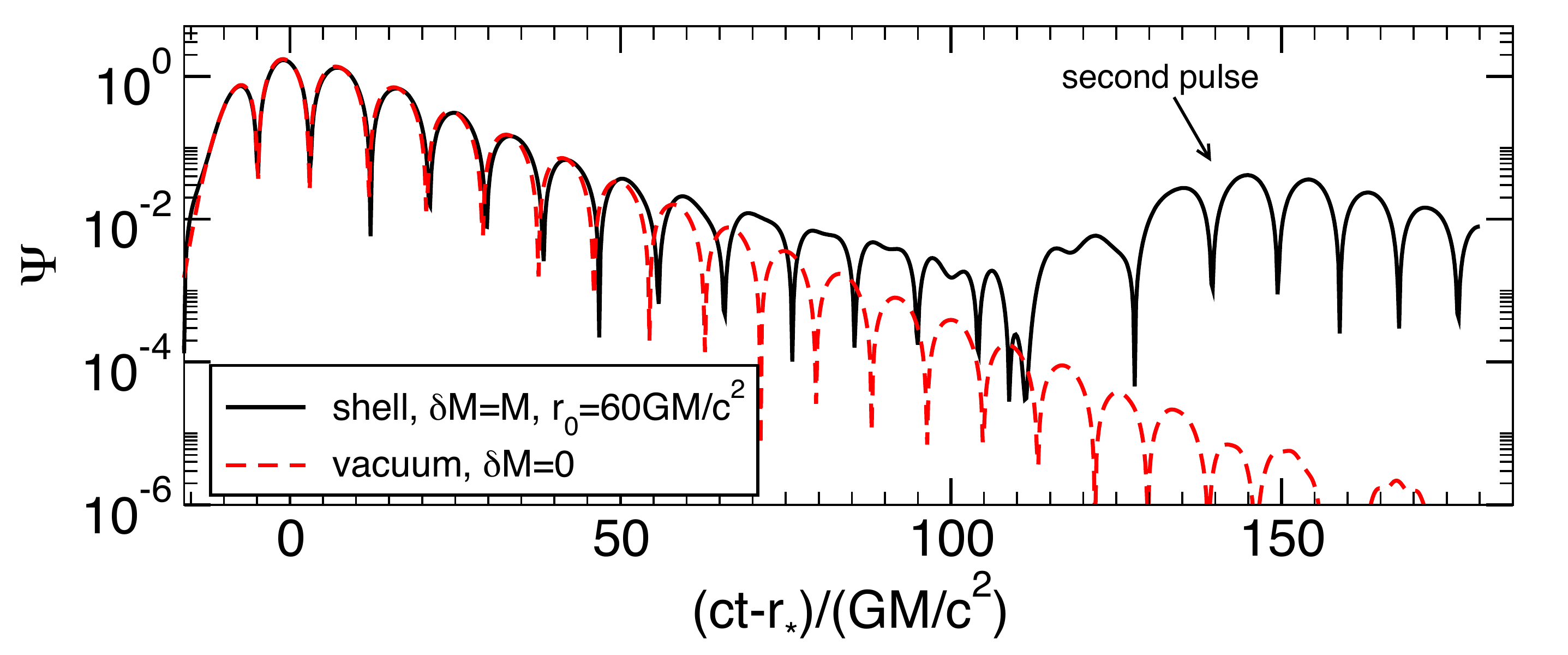}
\caption{
 Axial waveform for a Gaussian pulse, scattered off a system comprised of a Schwarzschild BH and a thin shell located at $r_0=60 GM/c^2$ and with mass equal to the BH mass $M$. The Gaussian pulse is initially placed near the ISCO, with width $\sigma=4GM/c^2$. 
 Note that the signal is initially dominated by the isolated BH QNMs, with the effect of matter appearing as a second pulse
 at later times.
\label{fig:ringdown} Adapted from Ref.~\cite{Barausse:2014pra}.}
\end{figure}

A similar phenomenon can take place in the presence of matter far away from the BH, which can act as a partially reflective mirror for outgoing radiation. Indeed,
the matter's gravitational potential can give rise to  a ``bump'' in the potential $V$ appearing in Eq.~\eqref{psiteuk}~\cite{Barausse:2014tra,Barausse:2014pra}. In this case, after the initial ringdown ``burst'', one would have echoes delayed by multiples of twice the light travel time between the circular photon orbit and the far away matter. An example from Ref.~\cite{Barausse:2014pra} is shown in Fig.~\ref{fig:ringdown}.
As can be seen, the initial ringdown transient is governed by the QNM frequencies and decay times of the pure Kerr/Schwarzschild system (without any matter), while only after a delay of 
twice the light travel time between the circular photon orbit and the matter
does the signal start to be described by the QNMs of the composite system. The same applies
to the case of a BH mimicker with near horizon reflectivity~\cite{Barausse:2014tra,Barausse:2014pra,Cardoso:2017cfl}: the initial ringdown transient
has spectrum comprised of the Kerr/Schwarzschild QNMs, while the ``true'' QNMs (accounting
for the modified boundary conditions at the horizon) only become apparent 
after a time roughly equal to twice
the light travel time between the near horizon mirror and the potential's peak.

The presence of echoes in the post-ringdown signal of some of the LIGO-Virgo systems
has been put forward by Ref.~\cite{Abedi:2017isz,Abedi:2018npz,Abedi:2018pst}, but the claim is highly controversial as later analyses have found it to have  low significance~\cite{Westerweck:2017hus,Nielsen:2018lkf}. Such a low significance (if any) is not surprising since the systems detected by LIGO-Virgo have very small ringdown SNR. Nevertheless, third generation ground-based detectors and LISA will allow for measuring the ringdown phase with much higher SNR~\cite{ringdown_letter}, which could open the way to detecting echoes, if there are any.

A more robust way to assess the possible presence of echoes is to look at their interaction with BH spins.
Energy conservation dictates that the echoes should have successively decreasing amplitude
around a non-spinning BH, but their amplitude can actually increase around BHs with sufficiently large spins.
This is due to a phenomenon called super-radiance~\cite{Penrose:1969pc,zeldovich1,zeldovich2,Brito:2015oca}, which is strictly related 
to the existence of an ergoregion (i.e. a near-horizon region where modes with
negative energy can exist as seen by an observer at infinity), and whereby echoes can extract rotational energy from the spinning BH. Indeed, BHs with sufficiently large spins
and surrounded by a ``mirror''  at the horizon become linearly unstable because of super-radiance, meaning
that the QNMs of the system are not damped but exponentially growing (at the expense of the BH's rotational energy).
This instability is also known as ``BH bomb'' instability~\cite{Press:1972zz,Cardoso:2004nk}, and if present would have important consequences for
the background of unresolved GWs that LIGO and Virgo are trying to detect~\cite{barausse:2018vdb}. Indeed, if all
astrophysical BH candidates in the universe (in binaries or isolated) had a perfectly reflective mirror at the horizon,
or if they had no horizon at all and they let gravitational radiation go through them undamped,
under reasonable assumptions on the distributions of their spins, their exponentially growing QNMs would  sum up to create
a stochastic background of GWs orders of magnitude larger than the current upper bounds from LIGO-Virgo~\cite{barausse:2018vdb}. This 
allows us to exclude that all BH candidates in the universe have a perfectly reflective horizon surface,
or that they let gravitational radiation go through them undamped. Indeed, 
LIGO-Virgo at design sensitivity will only allow for 1\% or less of the BH candidate population to have such properties.
The constraint can be extended to partially reflective mirrors or to objects that damp passing radiation. However, a transmissivity -- or energy loss 
through the object -- of at least 60\% (6\%) can
quench the instability for any spin (for spins $a\lesssim 0.9$), thus evading these constraints~\cite{barausse:2018vdb,Maggio:2018ivz}.

Let us note that a similar BH bomb super-radiant
instability affects BHs with sufficiently large spins and a mirror at large distances from the BH~\cite{Press:1972zz,Cardoso:2004nk}. Indeed,
if this mirror has sufficiently large reflectivity, it can focus gravitational radiation back to
the BH. This reflected radiation would in turn be scattered back by the effective potential of Eq.~\eqref{psiteuk},
and so on. Since the ergoregion of Kerr BHs extends (on the equatorial plane) to $r=2 r_s$ in Boyer-Lindquist coordinates, while the circular photon
orbit has radius that goes to $r_s$ in the same coordinates for $a\to 1$~\cite{bardeen}, it is clear that while bouncing
back and forth between the potential's peak and the mirror the wave can get amplified by producing
negative energy ergoregion modes, as in the case of a mirror at the horizon. 

While matter around the BH will not typically provide sufficient reflectivity for this instability to pick up,
a physical implementation of this idea is provided by ultralight bosonic degrees of freedom. Indeed, 
such boson fields (be them scalar, vectors or tensors) can become super-radiance unstable around BHs with sufficiently large spin,
provided that their Compton wavelength is comparable to the BH horizon radius~\cite{Detweiler:1980uk}. The instability is akin to the
BH bomb with an external mirror, where the mirror is provided here by the mass, which 
constrains the boson field near the BH by allowing for bosonic bound states.
As a result of the instability, the BH spin quickly decreases, and the angular momentum is transferred to
the boson field, which evolves into a rotating dipolar condensate. This condensate is in turn expected 
to emit almost monochromatic GWs, as can be understood simply from the quadrupole formula~\cite{Yoshino:2013ofa,Brito:2014wla,East:2017ovw,East:2017mrj,Pani:2012vp,Pani:2012bp,Baryakhtar:2017ngi}. Summing up the monochromatic waves that would be produced by all the (isolated) BHs in the universe, under reasonable assumptions for their spins, one
obtains a stochastic background of GWs that exceeds the current LIGO-Virgo bounds by several orders of 
magnitude if the boson's mass is in the right range~\cite{Brito:2017zvb,Brito:2017wnc}. In more details, current LIGO-Virgo bounds
on the stochastic background~\cite{TheLIGOScientific:2016wyq}
exclude boson masses between roughly   $2\times10^{-13}$ eV and $10^{-12}$ eV~\cite{Brito:2017zvb,Brito:2017wnc}. Similarly, LISA will allow
for ruling out masses in the range  $\sim 5\times[10^{-19},10^{-16}]$ eV~\cite{Brito:2017zvb,Brito:2017wnc}. Further constraints may 
come from trying to detect the monochromatic GWs produced by the bosonic condensate directly, either
in isolated BHs, or in the remnant resulting from a BH merger~\cite{Brito:2017zvb,Brito:2017wnc,Arvanitaki:2010sy,Brito:2014wla,Arvanitaki:2014wva,Arvanitaki:2016qwi,Baryakhtar:2017ngi,Dev:2016hxv}.

\section{Conclusions}\label{PRD}
In this pedagogical note, I have tried to show how the GW signal from the merger
of two BHs can be understood qualitatively based on very simple physical ingredients, including
the quadrupole formula, which allows for understanding the low frequency inspiral
and its dependence on the chirp mass; the PN precession of the spins during the inspiral;
the presence of an effective ISCO, whose position depends on the BH spins 
as a result of  ``frame dragging'' and which affects the overall power emitted in GWs; and the circular photon orbit and its effective potential,
whose physics determines the QNMs. I have also shown how modifications of GR and/or 
of the BH paradigm can affect each of these ingredients and produce GW signals differing from
GR. In particular, I have briefly outlined low frequency modifications to the gravitational waveform
that may arise from BH dipole radiation (caused in turn by BH hairs); possible changes to the conservative dynamics, which may result in changes of the ISCO properties and therefore in a different final spin; changes in the physics of the circular photon orbit, which would affect the properties of QNMs; and changes in the boundary conditions satisfied by 
the QNMs at the horizon and at infinity (as a result of exotic near horizon physics, matter far from the binary, or the presence of an ultralight boson field), which would result in echoes and super-radiant instabilities.

\begin{acknowledgments}
This work has received funding from the European Research Council (ERC) under the European Union's Horizon 2020 research and innovation programme (grant agreement no.~GRAMS-815673; project title ``GRavity from Astrophysical to Microscopic Scales'').
This work was also supported by the H2020-MSCA-RISE-2015 Grant No.~StronGrHEP-690904. The author would like to acknowledge networking support by the COST
Action CA16104.
\end{acknowledgments}

\bibliographystyle{apsrev4-1}

\bibliography{merged2}
\end{document}